\documentclass{iopjournal}
\usepackage{amssymb}

\usepackage{amsmath}
\input{Qcircuit}

\begin{document}

\articletype{Paper} 

\title{A High-Dimensional Quantum Blockchain Protocol Based on Time- Entanglement}

\author{Arzu Akta\c{s}$^{1,*}$ \orcid{0000-0001-9571-8012} and  \.Ihsan Y\i lmaz$^2$\orcid{0000-0001-7684-9690} }

\affil{$^1$Department of Statistics, Faculty of Science, \c{C}anakkale Onsekiz Mart University, \c{C}anakkale, Turkiye, 17020}

\affil{$^2$Department of Basic Sciences, Faculty of Engineering and Natural Sciences, Maltepe University, İstanbul, Turkiye, 34857}

\affil{$^*$Author to whom any correspondence should be addressed.}

\email{arzu.aktas@comu.edu.tr}


\begin{abstract}
Rapid advancements in quantum computing and machine learning threaten the long-term security of classical blockchain systems, whose protection mechanisms largely rely on computational difficulties. In this study, we propose a quantum blockchain protocol whose protection mechanism is directly derived from quantum mechanical principles. The protocol combines high-dimensional Bell states, time-entanglement, entanglement switching, and high-dimensional superdense coding. Encoding classical block information into time-delimited qudit states allows block identity and data verification to be implemented through the causal sequencing of quantum measurements instead of cryptographic hash functions. High-dimensional coding increases the information capacity per quantum carrier and improves noise resistance. Time-entanglement provides distributed authentication, non-repudiation, and tamper detection across the blockchain. Each block derives its own public-private key pair directly from the observed quantum correlations by performing high-dimensional Bell state measurements in successive time steps. Because these keys are dependent on the time ordering of measurements, attempts to alter block data or disrupt the protocol's timing structure inevitably affect the reconstructed correlations and are revealed during validation. Recent advances in the creation and detection of high-dimensional time-slice entanglement demonstrate that the necessary quantum resources are compatible with emerging quantum communication platforms. Taken together, these considerations suggest that the proposed framework can be evaluated as a viable and scalable candidate for quantum-secure blockchain architectures in future quantum network environments.
\end{abstract}

\section{Introduction}

The increasing scale and complexity of digital information systems have intensified concerns regarding data integrity, authentication, and long-term security. Traditional cryptographic techniques, which underpin many current blockchain architectures, rely heavily on assumptions regarding the computational difficulty. Although these assumptions are effective against classical adversaries, the rapid advancement of quantum computing and machine learning raises fundamental questions about their sustainability. In particular, quantum algorithms, capable of efficiently solving problems that are classically unsolvable, are prompting a revision of security frameworks that are not solely dependent on computational difficulty.

Within the field of quantum cryptography, various protocols have been developed to address these concerns by leveraging the physical properties of quantum systems. Among these, high-dimensional quantum schemes have attracted increasing interest owing to their ability to encode more information per carrier and their enhanced resistance to noise. Experimental and theoretical studies have shown that high-dimensional entanglement can increase robustness against decoherence, improve communication capacity, and strengthen resistance to copy-based attacks \cite{fiber,ecker,Cozzolino}. These advantages make high-dimensional quantum resources particularly attractive for distributed applications where both scalability and security are critical.

In conventional two-dimensional quantum technologies, a single qubit is typically encoded per photon. By contrast, high-dimensional quantum systems enable the encoding of multiple qubits within a single photon, with the number of encoded qubits determined by the dimensionality of the Hilbert space. Luo \emph{et al.}  proposed a scheme for teleporting arbitrarily high-dimensional photonic quantum states and experimentally demonstrated the teleportation of a qutrit~\cite{luo}. Moreover, Bang has investigated the controlled teleportation of high-dimensional quantum states using generalized Bell-state measurements~\cite{bang}.

Quantum memory plays a crucial role in quantum information processing. While two-dimensional protocols often require substantial quantum memory resources, high-dimensional implementations can significantly reduce memory requirements when projected onto effective two-dimensional subspaces. In this context, Wu \emph{et al.} have experimentally demonstrated high-dimensional entanglement between a photon and a multilayer atomic quantum memory~\cite{li}. Similarly, Wang \emph{et al.} have studied efficient quantum memory for orbital angular momentum qubits in cold atomic ensembles~\cite{wang}.

Recent advances in high-dimensional quantum communication provide a strong foundation for rethinking blockchain security in the presence of quantum adversaries. In particular, cavity-filtered biphoton frequency combs have been shown to support extremely large Hilbert-space dimensionalities through coherent time--frequency encoding, enabling robust certification of high-dimensional entanglement via Schmidt mode decomposition and Franson interferometry~\cite{chang_648_2021}. Such high-dimensional time--frequency entanglement offers not only increased information capacity but also enhanced resistance to noise, loss, and forgery---properties that are critically important for secure blockchain architectures. Furthermore, recent demonstrations of high-dimensional quantum key distribution based on energy--time and time--frequency entanglement have achieved high photon information efficiency and long-distance robustness over optical fiber networks, highlighting their suitability for distributed and tamper-resistant ledger systems~\cite{Liu_2024, cheng_high-dimensional_2023, scarfe2025}. These developments indicate that high-dimensional entanglement constitutes a key enabling resource for quantum blockchain protocols, where security can be enforced by fundamental quantum correlations rather than computational assumptions.

In classical blockchain architectures, blocks are sequentially connected through cryptographic hash values and time-stamping mechanisms. This structure guarantees that altering the content of any individual block breaks the integrity of all blocks that follow it. Nevertheless, the advent of quantum computing exposes fundamental weaknesses in such classically secured blockchain protocols, indicating that purely classical security techniques may no longer be sufficient against quantum-enabled attacks. To address this issue, Rajan et al. have proposed a quantum blockchain protocol that makes use of time entanglement~\cite{rajan}. In related work, Caram`es et al. reviewed existing blockchain cryptographic techniques and discussed their resistance to attacks arising from quantum computing~\cite{carames}.

This work builds on our earlier study of high-dimensional quantum digital signatures based on entanglement swapping~\cite{arzu}, in which multi-party authentication was achieved using spatially entangled qudit states.

In this paper, we present a quantum blockchain framework that relies on time entanglement in high-dimensional Bell states (HDBS) to improve both data capacity and security. By applying high-dimensional Bell-state measurements (HDBM) to the blockchain blocks, public and private cryptographic keys are generated in a manner that is directly tied to the underlying quantum states. This approach provides a scalable and robust basis for blockchain architectures secured through quantum mechanical principles.

\section{Blockchain and Security Model}

This section outlines the blockchain model adopted in this work and clarifies
the security assumptions on which the proposed protocol is based. In general
terms, a blockchain consists of an ordered collection of blocks that together
form a distributed ledger, with each block storing data as well as information
that links it to its predecessor. 

In conventional blockchain architectures, this linkage is typically realized through cryptographic hash functions combined with time-stamping mechanisms, which are designed to support data integrity and traceability throughout the ledger. The security of such systems is largely based on assumptions of computational hardness, realized through hash functions, digital signatures, and consensus mechanisms. However, these assumptions become increasingly fragile in the presence of large-scale quantum computers, making classical blockchain protocols vulnerable to quantum-enabled attacks.

In contrast, a quantum blockchain is considered in this work as a distributed ledger whose security is ensured by quantum resources rather than computational assumptions. Each block is associated with quantum states and measurement outcomes that are used for authentication and data integrity verification. Instead of relying on cryptographic hash functions, the proposed quantum blockchain establishes cryptographic links between blocks using high-dimensional time entanglement.

Within this framework, we consider a blockchain consisting of $n$ consecutive blocks, denoted by $B_1, B_2, \dots, B_n$. Each block stores classical data and interacts with other blocks through a quantum communication protocol. Blocks may behave honestly or adversarially, and may attempt to alter data, impersonate other blocks, or distribute false information. Communication between blocks is supported by quantum channels capable of distributing high-dimensional, time-entangled quantum states, along with authenticated classical communication channels.

The proposed protocol is designed to achieve the following security objectives:
\begin{itemize}
	\item \textbf{Data integrity:} Any unauthorized modification of block data during transmission or storage must be detectable.
	\item \textbf{Authentication:} Blocks must be able to verify the origin of received data.
	\item \textbf{Non-repudiation:} A block that generates data must not be able to deny its participation at a later stage.
	\item \textbf{Forgery resistance:} Adversarial blocks must not be able to create or alter data that successfully passes the verification process.
	\item \textbf{Distributed verification:} Verification is carried out collectively by multiple blocks, without relying on a single trusted authority.
\end{itemize}

Within this setting, the objective of this study is to design a quantum blockchain protocol that leverages high-dimensional quantum states and time entanglement to provide security guarantees that are traditionally achieved through computational hardness assumptions in classical blockchains. By exploiting the advantages of high dimensionality and time entanglement, the proposed protocol enables distributed and verifiable block authentication, secure data transmission, and tamper detection through high-dimensional Bell-state measurements and entanglement exchange over time.

\section{Preliminaries}

In quantum information processing, all physical operations other than measurement are described by unitary transformations. A linear operator $U$ acting on a Hilbert space is said to be unitary if it satisfies
\begin{equation}\label{ee1}
	U^\dagger = U^{-1},
\end{equation}
where $U^\dagger = (U^*)^T$ denotes the Hermitian conjugate of $U$.

For qubit systems ($N=2$), common examples of unitary operators include the Hadamard ($H$) and Pauli ($X,Y,Z$) gates. Their generalizations to $N$-dimensional systems (qudits) form the basis of high-dimensional quantum information processing and are essential for constructing multi-qudit quantum gates.

\vspace{0.3cm}
\noindent
\textbf{High-dimensional cat states:}
An $n$-particle high-dimensional cat (GHZ-type) state in an $N$-dimensional Hilbert space is defined as~\cite{zhaoDef,dboyut}
\begin{equation}\label{ee6}
	\ket{\psi(x_1,\dots,x_n)} =
	\frac{1}{\sqrt{N}}
	\sum_{j=0}^{N-1}
	w^{j x_1}
	\ket{
		j,\;
		j+x_2 \!\!\mod N,\;
		\dots,\;
		j+x_n \!\!\mod N
	},
\end{equation}
where $x_1,\dots,x_n \in \{0,1,\dots,N-1\}$ and
\begin{equation}
	w = e^{\frac{2\pi i}{N}}.
\end{equation}
These states represent maximally entangled multipartite qudit systems and play a key role in high-dimensional quantum communication protocols.

\vspace{0.3cm}
\noindent
\textbf{High-dimensional Bell states:}
The $N$-dimensional entangled Bell states for a two-qudit system are given by~\cite{zhaoDef,frontiers}
\begin{equation}\label{ee2}
	\ket{\psi(x,y)} =
	\frac{1}{\sqrt{N}}
	\sum_{j=0}^{N-1}
	w^{j x}
	\ket{j} \otimes \ket{j+y \!\!\mod N},
\end{equation}
where $x,y \in \{0,1,\dots,N-1\}$.  
The set $\{\ket{\psi(x,y)}\}_{x,y=0}^{N-1}$ forms a complete orthonormal Bell basis in the composite Hilbert space $\mathbb{C}^N \otimes \mathbb{C}^N$.

\vspace{0.3cm}
\noindent
\textbf{Bell-state transformation operator:}
The unitary operator that maps the computational basis onto the high-dimensional Bell basis is defined as
\begin{equation}\label{ee4}
	U_{(x,y)} =
	\sum_{j=0}^{N-1}
	w^{xj}
	\ket{j+y \!\!\mod N}\bra{j}.
\end{equation}
This operator satisfies
\begin{equation}
	U_{(x,y)}^\dagger U_{(x,y)} = I,
\end{equation}
and therefore constitutes a valid unitary transformation. Such operators form the core of high-dimensional Bell-state measurements, which are fundamental for quantum teleportation, entanglement swapping, and quantum blockchain protocols based on high-dimensional Bell states.

\section{Encoding Classical Blocks into Time-Entangled Bell States}
The n-block protocol will be explained step-by-step. However, to make the protocol more understandable, some information that will be used in these steps is explained below first.
\vspace{0,3cm}

The equation (\ref{bb1}) encodes the 2-bit information of a classical block into a time Bell state created between photons absorbed at two different times. For a two-bit classical register $(b_1,b_2)\in\{0,1\}^2={(0,0),(0,1),(1,0),(1,1)}$, where $b_1,b_2 \in \{0,1\}$ represent binary variables, the corresponding time-bin Bell state generated at times $t=0$ and $t=\tau$ is given by~\cite{rajan,Gao}.
\begin{equation}\label{bb1}
	\ket{B_{b_1 b_2}}^{(0,\tau)} =
	\frac{1}{\sqrt{2}}
	\left(
	\ket{0^{0}}\ket{b_2^{\tau}}
	+
	(-1)^{b_1}
	\ket{1^{0}}\ket{\overline{b_2}^{\tau}}
	\right),
\end{equation}
where the superscripts indicate the time bins associated with the photonic modes, and $\overline{b_2}=b_2\oplus 1$ denotes the logical complement.

Equation (\ref{bb1}) describes the two-dimensional case in which the classical two-bit record of a block is encoded into a time-entangled Bell state shared between two absorption times. While this representation is sufficient for illustrating time-entanglement, it limits both the amount of information that can be encoded in a single block and the robustness of the protocol against noise.

To overcome these limitations, the construction in (\ref{bb1}) is extended to higher dimensions by replacing qubit states with qudit states. In this generalization, the binary basis \{0,1\} is replaced by an N-dimensional computational basis $\{0,1,\ldots,N-1\}$, and the relative phase factor $(-1)^{b_1} $is generalized to the N-th root of unity $\omega = e^{2\pi i/N}$.

As a result, the time-entangled Bell state takes the high-dimensional form given in (\ref{b1a}), where the classical record $(b_1,b_2)$  is encoded into correlations between qudit states absorbed at two different times. This generalization allows each block to carry a larger amount of information while preserving the time-entanglement structure required for linking blocks in the quantum blockchain.
To enhance both information capacity and security, the classical two-bit register is generalized to an $N$-dimensional alphabet, where $b_1,b_2 \in \mathbb{Z}_N = \{0,1,2,\dots,N-1\}$ denote integers defined modulo $N$, and $w = e^{\frac{2\pi i}{N}}$. Each classical block is then encoded into a high-dimensional Bell state entangled across two distinct time bins:
\begin{equation}\label{b1a}
	\ket{HDBS_{b_1 b_2}}^{(0,t)}
	=
	\ket{\beta_{b_1 b_2}}^{(0,t)}
	=
	\frac{1}{\sqrt{N}}
	\sum_{j=0}^{N-1}
	w^{j b_2}
	\ket{j^{0}}
	\ket{(b_1+j)\!\!\mod N^{\,t}}.
\end{equation}
This construction yields a set of $N^2$ orthonormal high-dimensional Bell states distributed over time, forming the fundamental building blocks of the proposed quantum blockchain.
\vspace{0.3cm}
\noindent
\section{High-Dimensional Bell type Quantum Blockchain Protocol with Entanglement in Time}

Since it is possible to experimentally realize Bell states with current technologies \cite{PRA2025,cheng_high-dimensional_2023,chang_648_2021}, this study investigates high-dimensional time-entangled Bell states. 

In this section, we present a high-dimensional quantum blockchain protocol based on time-entanglement.

\begin{figure}[h!]
	\centering
	\includegraphics[width=0.8\textwidth]{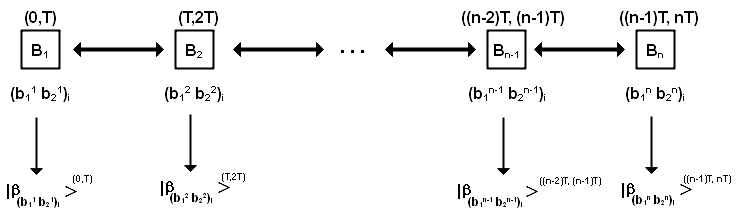}
	\caption{High-dimensional quantum blockchain scheme based on time-entanglement for an arbitrary number of blocks.}
	\label{res4}
\end{figure}

More generally, let $B_1, B_2, \dots, B_n$ represent the blocks of a classical blockchain. Each block initially contains a classical register encoding two symbols. These classical registers are mapped onto time-entangled quantum states by associating the logical information of each block with photons generated (or detected) at distinct time instants. To further increase the information density of each block, the classical register $(b_1 b_2)$ is extended to a sequence $(b_1 b_2)_i$, where $i=1,2,\dots,m$. Each pair $(b_1 b_2)_i$ is independently encoded into a high-dimensional time-entangled Bell state, resulting in a composite quantum block. The overall $n$-block high-dimensional quantum blockchain architecture is illustrated in Fig.~\ref{res4}.

The proposed quantum blockchain protocol consists of two main phases. These phases are:
\begin{itemize}
	\item \textbf{Key generation and sharing stage}, where high-dimensional Bell-state measurements are employed to establish shared quantum keys between authorized parties.
	\item \textbf{Messaging and verification stage}, in which block integrity and authenticity are verified using the correlations imposed by time-entanglement.
\end{itemize}
\vspace{0.3cm}

\subsection { \textbf{Key Generation and Sharing Step} }
\begin{enumerate}
	\item Consider the transfer of classical data stored in block $B_1$  to block $B_n$.  
	The classical data contained in block $B_1$ is represented as an $m$-symbol string
	\begin{equation}\label{bb2}
		\mathbf{d}^{(1)} =
		\left(
		data_1^{(1)},\;
		data_2^{(1)},\;
		\dots,\;
		data_m^{(1)}
		\right),
	\end{equation}
	where each symbol satisfies $data_i^{(1)} \in \mathbb{Z}_N$.
	\vspace{0.2cm}
	The corresponding quantum representation of this data, encoded in the computational basis of $N$-dimensional qudits, is given by
	\begin{equation}\label{bb3}
		\ket{d_{B_1}}
		=
		\bigotimes_{i=1}^{m}
		\ket{data_i^{(1)}}
		=
		\ket{data_1^{(1)}}\ket{data_2^{(1)}}\cdots\ket{data_m^{(1)}}.
	\end{equation}
	\vspace{0.2cm}
	To enhance protocol security against eavesdropping and forgery, the data qudits are transformed into a mutually unbiased basis. This transformation is realized by applying a unitary operator $U$ to each qudit, where $U$ maps the computational basis $\{\ket{j}\}$ onto a new basis $\{\ket{\xi_j}\}$ such that
	\begin{equation}
		U\ket{j} = \ket{\xi_j}, \qquad j \in \mathbb{Z}_N.
	\end{equation}
	\vspace{0.2cm}
	After the basis transformation, the encoded quantum state associated with block $B_1$ becomes
	\begin{equation}\label{bb4}
		\ket{ID_{B_1}}
		=
		\bigotimes_{i=1}^{m}
		U\ket{data_i^{(1)}}
		=
		\bigotimes_{i=1}^{m}
		\ket{\xi_{data_i^{(1)}}},
	\end{equation}
	where $\ket{ID_{B_1}}$ denotes the encoded quantum identifier of block $B_1$, which will be used in the subsequent key generation and sharing process.
	\vspace{0.2cm}
	\item For a blockchain consisting of $n$ blocks, a high-dimensional Bell-state measurement is performed in block $B_{n-1}$ at time $t=(n-1)T$, thereby creating entanglement between photons generated at times $t=(n-2)T$ and $t=nT$. By iteratively performing HDBM in intermediate blocks at times $(n-2)T,(n-3)T,\dots,T$, an entanglement-swapping process is established that ultimately generates time-entanglement between photons associated with blocks $B_1$ and $B_n$.
	
	This sequence of time-entanglement ordered measurements creates a time-entangled quantum channel linking the first and last blocks of the blockchain. Simultaneously, each HDBM produces verifiable classical records that encode when and how data are generated, transmitted, and modified. Based on these measurement outcomes, each block autonomously generates its own public--private key pair $(B_p,B_s)$, enabling authentication, data integrity verification, and detection of any external intervention within the blockchain.
	
	\vspace{0.2cm}
	\item For a quantum blockchain consisting of $n$ blocks, the high-dimensional Bell-state measurement performed in block $B_{n-1}$ yields an outcome
	\begin{equation}
		(B_{n-1})_s =	\bigotimes_{i=1}^{m}((B_{n-1}^1)_i \  (B_{n-1}^2)_i)=
		\left(
		(B_{n-1}^1)_1\,(B_{n-1}^2)_1\ ,
		\dots\ ,
		(B_{n-1}^1)_m\,(B_{n-1}^2)_m\ 
		\right)
	\end{equation}
	while the corresponding public key is given by
	\begin{equation}
		(B_{n-1})_p =  \oplus _{i=1}^{m}((B_{n-1}^1)_i \  (B_{n-1}^2)_i)=
		\left(
		(B_{n-1}^1)_1 \oplus (B_{n-1}^2)_1,\;
		\dots,\;
		(B_{n-1}^1)_m \oplus (B_{n-1}^2)_m
		\right)
	\end{equation}
	respectively.
	\vspace{0.2cm}
	The private key $(B_{n-1})_s$ is stored locally in block $B_{n-1}$, while the public key $(B_{n-1})_p$ is distributed to all other blocks except $B_1$ via high-dimensional superdense coding. Each receiving block performs a local high-dimensional Bell-state measurement to reconstruct and store the authentic public key $(B_{n-1})_p$, thereby enabling verification and traceability across the blockchain.
	\vspace{0.2cm}
	
	\item For a quantum blockchain consisting of $n$ blocks, each intermediate block $B_k$ $(2 \leq k \leq n-1)$ performs a high-dimensional Bell-state measurement at time $t=kT$. The corresponding measurement outcome
	\begin{equation}
		(B_k^1, B_k^2) \in \mathbb{Z}_N \times \mathbb{Z}_N
	\end{equation}
	is used to generate a private--public key pair $(B_k)_s$ and $(B_k)_p$ according to the same construction described above.
	
	The private key $(B_k)_s$ is stored locally within block $B_k$, while the public key $(B_k)_p$ is distributed to all other blocks (except $B_1$) using high-dimensional superdense coding. Through the sequence of time-entanglement ordered Bell-state measurements, a global time-entangled structure is formed across the blockchain, enabling authentication, traceability, and detection of any unauthorized modification of block data.
	\vspace{0.2cm}
	\item In summary, each block in the quantum blockchain independently generates a private--public key pair based on the outcomes of its local high-dimensional Bell-state measurements. The private keys are kept locally in their respective blocks, whereas the associated public keys—except for block $B_1$—are shared with the remaining blocks through high-dimensional superdense coding. In this way, quantum security requirements are maintained while still allowing verification and authentication to be carried out across the entire blockchain.
	\vspace{0.2cm}
	
	\item For a quantum blockchain consisting of $n$ blocks, the same procedure applies. Following the HDBM performed in block $B_1$ at time $t=T$, the public key $(B_1)_p$ is shared exclusively with block $B_n$ through the time-entangled channel linking $B_1$ and $B_n$. As a result, the quantum state reconstructed in block $B_n$ is given by
	\begin{equation}\label{b5}
		\ket{d_{B_n}} =
		\bigotimes_{i=1}^{m}
		U_{j_i k_i}^\dagger
		\ket{d_{B_1}},
	\end{equation}
	where
	\begin{equation}
		j_i = \bigoplus_{r=1}^{n-1} (B_r)_{p_i^1},
		\qquad
		k_i = \bigoplus_{r=1}^{n-1} (B_r)_{p_i^2},
		\qquad i=1,2,\dots,m.
	\end{equation}
	
	In addition, block $B_1$ obtains a time-stamped global identity as a result of the measurement performed in the mutually unbiased basis at time $t=T$. This identity is defined as
	\begin{equation}\label{bb6}
		ID_{B_1}^{G}
		=
		\bigotimes_{i=1}^{m}
		U_{(B_1)_{p_i^1}\,(B_1)_{p_i^2}}^\dagger
		\ket{\xi_{data_i^{(1)}}},
	\end{equation}
	and is published as the public identity of block $B_1$. The time-stamped global identity $ID_{B_1}^{G}$ is obtained as in Eq.~(\ref{bb6}) and published as the public identity of block $B_1$, enabling global verification and non-repudiation across the quantum blockchain.
	\vspace{0.2 cm}
	\item By performing measurements on new bases in equation (\ref{b5}) in block $B_n$, the identity of block $B_1$ is calculated as $ID_{B_n}^{B_1}$. Here, the subscript and superscript refer to the block to which the data will be transferred and the block that actually owns the identity, respectively.
	
\end{enumerate}

\subsection{{\textbf{Messaging and Validation Step}}}

\begin{enumerate}
	
	\item The same messaging and validation procedure applies to a quantum blockchain consisting of $n$ blocks. The quantum data $\{data_i^{(1)}\}$ is teleported from block $B_1$ to block $B_n$, while the public key $(B_1)_p$ is available to the receiving block. In the presence of adversarial modification, the altered pair $\{\overline{data}_i^{(1)}, (\overline{B_1})_p\}$ is obtained.
	
	\begin{enumerate}[(a)]
		
		\item \textbf{Validation--1 (Data integrity check):}  
		In block $B_n$, the altered global identity $\overline{ID}_{B_1}^G$ is computed using Eq.~(\ref{bb6}) and compared with the published identity:
		\begin{equation}
			(\overline{ID}_{B_1}^G)_i = (ID_{B_1}^G)_i,
			\qquad i=1,2,\dots,m.
		\end{equation}
		
		\item \textbf{Validation--2 (Identity authenticity check):}  
		The reconstructed identity $ID_{B_n}^{B_1}$ obtained via measurements in the mutually unbiased basis is compared with the published identity $ID_{B_1}^G$:
		\begin{equation}
			i=1,\dots,m \quad
			\begin{cases}
				(ID_{B_1}^G)_i = (ID_{B_n}^{B_1})_i,
				& \text{if } \displaystyle \bigoplus_{r=2}^{n-1} (B_r)_{p_i} = 0, \\[6pt]
				(ID_{B_1}^G)_i \neq (ID_{B_n}^{B_1})_i,
				& \text{if } \displaystyle \bigoplus_{r=2}^{n-1} (B_r)_{p_i} \neq 0.
			\end{cases}
		\end{equation}
		
	\end{enumerate}
	
	\item
	If the data successfully passes all validation checks performed in block $B_n$, the verified triple
	\begin{equation}
		\{\overline{data}_i^{(1)},\,(\overline{B_1})_p,\,\overline{ID}_{B_n}^{B_1}\}
	\end{equation}
	is transmitted to the remaining blocks $B_l$, where $1<l<n$. Each receiving block independently executes the same validation steps to assess data authenticity and integrity. This distributed verification mechanism ensures that any malicious modification is detected and that only valid data is propagated throughout the blockchain.

	\item For a quantum blockchain consisting of $n$ blocks, identity verification is again performed solely with respect to the public global identity $ID_{B_1}^G$ published by block $B_1$. Any identity information forwarded from block $B_n$ is not treated as authoritative.
	
	\vspace{0.2cm}
	In an intermediate block $B_l$ $(1<l<n)$, the identity associated with block $B_n$ is reconstructed as
	\begin{equation}
		ID_{B_l}^{B_n}
		=
		\bigotimes_{i=1}^{m}
		U_{j_i k_i}^\dagger
		\left( \ket{ID_{B_1}^G} \right)_i ,
	\end{equation}
	where
	\begin{equation}
		j_i = \bigoplus_{\substack{r=2 \\ r \neq l}}^{n-1} (B_r)_{p_i^1},
		\qquad
		k_i = \bigoplus_{\substack{r=2 \\ r \neq l}}^{n-1} (B_r)_{p_i^2},
		\qquad i=1,2,\dots,m.
	\end{equation}
	
	The reconstructed identity $ID_{B_l}^{B_n}$ is then compared with the identity value forwarded from block $B_n$:
	\begin{equation}
		i=1,\dots,m \quad
		\begin{cases}
			(ID_{B_l}^{B_n})_i = (\overline{ID}_{B_l}^{B_1})_i,
			& \text{if } \displaystyle\bigoplus_{\substack{r=2 \\ r \neq l}}^{n-1} (B_r)_{p_i} = 0, \\[6pt]
			(ID_{B_l}^{B_n})_i \neq (\overline{ID}_{B_l}^{B_1})_i,
			& \text{if } \displaystyle\bigoplus_{\substack{r=2 \\ r \neq l}}^{n-1} (B_r)_{p_i} \neq 0 .
		\end{cases}
	\end{equation}
	This procedure allows block $B_l$ to determine whether any fraudulent activity has occurred in block $B_n$.
	\vspace{0.2cm}
	
	\item For a quantum blockchain consisting of $n$ blocks, the same procedure applies. In any intermediate block $B_h$ $(1<h<n)$, the \textbf{Validation--1} and \textbf{Validation--2} steps are executed exactly as in block $B_n$. Based on the outcomes of these checks, block $B_h$ determines whether the data originating from block $B_1$ is accepted or rejected. If the validation is successful, the verified triplet is propagated to the next block in the chain, thereby enabling sequential and distributed authentication across the quantum blockchain.
	\vspace{0.2cm}
	
	\item For a quantum blockchain consisting of $n$ blocks, if the data is successfully validated in block $B_n$, the verified triplet
	\begin{equation}
		\{\overline{data}_i^{(1)},\,(\overline{B_1})_p,\,\overline{ID}_{B_n}^{B_1}\}
	\end{equation}
	is forwarded to block $B_{n-1}$. The same validation steps are then executed in block $B_{n-1}$ as in any intermediate block $B_l$ $(1<l<n)$.
	
	If block $B_{n-1}$ accepts the data as authentic, the newly reconstructed identity $\overline{ID}_{B_{n-1}}^{B_n}$ is forwarded to block $B_{n-2}$ in the form
	\begin{equation}
		\{\overline{data}_i^{(1)},\,(\overline{B_1})_p,\,\overline{ID}_{B_{n-1}}^{B_n}\}.
	\end{equation}
	This sequential validation process continues through blocks $B_{n-2}, B_{n-3}, \dots, B_2$.
	
	\vspace{0.2cm}
	In this manner, each block performs verification using a reduced set of keys while still being able to detect any forgery or tampering occurring in preceding blocks. The sequential structure significantly improves scalability while preserving the security guarantees of the quantum blockchain protocol.
	
	To better illustrate the protocol we propose, the following section examines an example of a four-block quantum blockchain.
\end{enumerate}
\vspace{0.2cm}

\subsection{{\textbf{Illustrative Example: Four-Block Quantum Blockchain}}}
\begin{figure}[h!]
	\centering
	\includegraphics[width=0.8\textwidth]{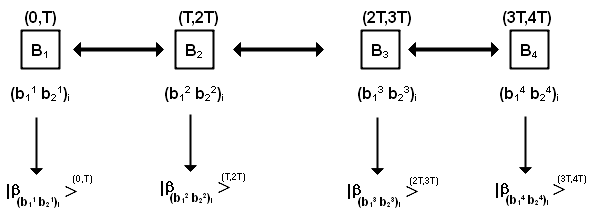}
	\caption{Illustration of a four-block high-dimensional quantum blockchain scheme based on time-entanglement.}
	\label{res3}
\end{figure}

Let $B_1, B_2, B_3, B_4$ denote four consecutive blocks in a classical blockchain, as illustrated in Fig.~\ref{res3}.

To further illustrate the operation of the protocol proposed in this section, we will consider an example consisting of four blocks, denoted by $B_1$, $B_2$, $B_3$, and $B_4$ respectively.

The $B_1$ block initially stores a classical data record $\{\text{data}^{(1)}_i\}_{i=1}^m$. Each data element is encoded into a high-dimensional time-entangled Bell state according to Equation (8), so that the resulting quantum state is distributed across two different time zones. This encoded state forms the quantum descriptor associated with $B_1$.

At the next time step $t = T$, the $B_2$ block performs a high-dimensional Bell state measurement on the locally available quantum systems. This measurement results in a quantum entanglement swap between the $B_1$ and $B_3$ blocks, thus extending time-entanglement across the blockchain. Based on the measurement result, the $B_2$ block generates a private-public key pair. The $B_2$ block stores the private key locally and distributes the corresponding public key to the other blocks (except the $B_1$ block) via high-dimensional superdense coding.

At $t = 2T$, the $B_3$ block also performs a high-dimensional Bell state measurement. This process further extends the time-entanglement structure, effectively linking the quantum states associated with $B_1$ and $B_4$. As in the previous step, the $B_3$ block creates its own private-to-public key pair, independent of the local measurement result.

Finally, the $B_4$ block receives the quantum data transmitted from $B_1$, along with the public keys generated by the intermediate blocks. By applying the validation procedures described in section 6.2, the $B_4$ block reconstructs the identity of $B_1$ and verifies both the integrity and authenticity of the data transmitted with this identity. Attempts to alter the data or disrupt the time sequence of entanglement switching operations lead to discrepancies detected during the validation process.

This four-block example demonstrates that the proposed protocol can be traced step-by-step for a finite blockchain. The proposed protocol has the infrastructure to maintain the same security features regardless of the number of blocks.

\section{Security Analysis}

The security of the quantum blockchain protocol presented in this work is based on the combined use of high-dimensional quantum states and time-entanglement. High-dimensional encoding makes it possible to transmit more information using a single quantum state and, at the same time, improves tolerance to noise. It also increases the protocol’s ability to withstand both internal and external attacks. Taken together, these features play an important role in enhancing the overall security of the proposed system.

The main feature that sets this work apart is the use of time-entanglement rather than spatial entanglement. With time-entanglement, quantum states are connected across different time intervals, and the security of the protocol depends on carrying out measurements in a well-defined causal order. If an adversary attempts to intercept, delay, or tamper with these time-entangled states, the normal behavior of the quantum states is disturbed due to the measurement rules of quantum mechanics, and this disturbance can be detected during the verification process.
The proposed protocol takes advantage of the physical limitation that strongly entangled quantum states cannot reliably form trusted quantum connections with unauthorized third parties at the same time. As a result, an attacker attempting to interfere with the system cannot extract useful information without disturbing the quantum structure shared by the legitimate blocks. Such disturbances naturally appear during the verification stage and make unauthorized access detectable. In a similar way, an eavesdropper trying to gain information about private or public keys through entanglement-based attacks inevitably reduces the reliability of the legitimate quantum states, thereby exposing the presence of the attack. This effect becomes even more pronounced when high-dimensional Bell states are used, making eavesdropping increasingly difficult as the system dimensionality grows.

The no-cloning theorem is a crucial security layer that prevents an adversary from copying quantum data and time-coupled keys. In the proposed protocol, identities and authentication keys are recovered through time-entangled high-dimensional Bell state measurements. Therefore, attempts to copy or duplicate these quantum elements will inevitably corrupt the underlying quantum states and be revealed during the distributed authentication process. This feature also limits the possibility of interfering with the protocol after the data has been generated.

Taken together, the use of high-dimensional quantum states and time-entanglement offers strong protection against forgery, identity impersonation, external attacks, and collusion between dishonest blocks. These properties make the proposed quantum blockchain protocol resilient to both classical and quantum adversaries, while still maintaining a level of scalability and robustness that is appropriate for future quantum network applications. 

Below, to support these claims, we will analyze the protocol's resilience to representative external and internal attack strategies.

\subsection*{Intercept--Resend Attack}

Consider an external adversary attempting an intercept--resend attack on the
time-entangled qudit states exchanged between two consecutive blocks.
Because the protocol relies on high-dimensional Bell correlations distributed across
distinct time bins, any interception necessarily involves a measurement on at least
one subsystem. Such a measurement irreversibly collapses the joint quantum state
and breaks the time correlations required for successful entanglement swapping.

As a consequence, the reconstructed public keys derived from high-dimensional
Bell-state measurements no longer satisfy the expected correlation relations, causing
both Validation--1 and Validation--2 to fail with high probability. The probability of
detecting such an attack increases with the Hilbert-space dimension $N$, rendering
intercept--resend strategies increasingly ineffective for large-dimensional systems.

\subsection*{Dishonest Block Collusion}

Assume that a subset of intermediate blocks colludes in an attempt to forge data
originating from block $B_1$. Successful forgery would require reconstructing the
correct global identity $ID^{G}_{B_1}$, which depends on measurement outcomes from all non-colluding blocks in the blockchain. Since private keys are generated locally and never shared, colluding blocks lack sufficient information to reproduce valid identity correlations.

As a result, any forged data inevitably fails the distributed validation process.
This ensures that the proposed protocol remains secure against internal collusion
attacks and preserves data integrity even in the presence of dishonest participants.

From the above analysis, several key security properties of the proposed protocol
follow, as listed below:

\begin{enumerate}
	\item \textbf{Non-repudiation:}  
	In the proposed protocol, the block that generates the data does not have access to the public keys of the other blocks involved in the validation process. Consequently, it cannot modify either the transmitted data or its own public key in a way that would allow it to successfully pass both Validation--1 and Validation--2. Since Validation--2 explicitly depends on the public keys of the intermediate blocks, any attempt by the sender to deny or manipulate the transmission is detected during the validation stage.
	
	\item \textbf{Transferability:}  
	In the proposed blockchain, data is only transmitted to subsequent blocks after authentication and validity conditions are successfully completed. The receiving block moves the data further down the chain after fulfilling the necessary verification requirements. If verification fails at any point, the transmission process is terminated at that stage. This prevents data from spreading across the network when identity or data verification is not possible.
	
	\item \textbf{Forgery resistance:}  
	Data altered or fabricated by the receiving block is detected during verification processes performed by sub-blocks upon transmission. Authentication in the proposed protocol relies on a global public identity generated by the original source block, not on information provided by intermediate receivers. Consequently, forged data cannot meet the necessary verification requirements and is detected before being accepted by the blockchain.
	
	\item \textbf{Message creation by the receiver:}  
	All the validation steps in the protocol rely on the publicly available identity of the source block. This prevents the receiving block from generating unauthorized messages on its own. Because the identity of the source block is independently verified by all blocks on the network, the receiving block cannot generate a message that successfully bypasses the validation process.
	
	\item \textbf{Message modification by the receiver:}  
	Any attempt by a receiving block to modify the transmitted data is detected during the verification stage. This is due to the fact that the protocol makes use of quantum-based information, where both the data and the corresponding global identifiers depend on inputs from multiple blocks. As a result, any inconsistent action introduced by a dishonest receiver disrupts the expected structure and becomes visible during authentication.
	
	\item \textbf{Internal attacks:}  
	The protocol employs a distributed verification mechanism in which blocks
	independently evaluate the integrity and authenticity of the received data.
	Through this process, a block can identify forged information or unauthorized
	modifications introduced at earlier stages of the blockchain. The sequential
	structure of the verification procedure limits the number of verification keys
	required at each step, thereby improving scalability while preserving the
	intended security guarantees.
	
	\item \textbf{External attacks:}  
	High-dimensional quantum states are secured against external threats by being coupled with time-entanglement. Interactions involving the hijacking, measurement, or manipulation of quantum states necessarily alter their correlational structure. Such deviations from expected behavior emerge during verification phases, enabling the detection of external interference.
	
\end{enumerate}

\section{Conclusion}

This work introduces a quantum blockchain protocol that makes use of
high-dimensional quantum states together with time-entanglement.
Within this framework, high-dimensional Bell states are employed in
combination with entanglement swapping and superdense coding to support
secure information transfer between blocks and to enable a distributed
authentication mechanism at the quantum level.

The use of high-dimensional quantum states offers clear advantages over low-dimensional (qubit-based) approaches, particularly in terms of information capacity and robustness against noise. In an N-dimensional Hilbert space, a single quantum carrier can encode up to $ \log_2 N $ qubits (or classical bits), leading to a logarithmic increase in information capacity with increasing dimensionality. Moreover, high-dimensional quantum states have been shown to exhibit improved tolerance to environmental noise and experimental imperfections. Recent experimental studies have demonstrated the generation and detection of high-dimensional time-bin entanglement in fiber-based systems, confirming that such resources are compatible with realistic quantum communication infrastructures and scalable optical platforms \cite{PRA2025,OE2023}.

A key contribution of the proposed protocol is the use of time-entanglement instead of spatial entanglement. Time entanglement enforces security through the causal ordering of quantum measurements, enabling uniform and distributed protection of all blocks in the blockchain. Attempts to intercept, delay, or manipulate time-entangled quantum states inevitably disturb the expected structure of the quantum system, resulting in inconsistencies that can be detected during the verification process. The practical relevance of this approach is further supported by recent experimental work on time-entangled quantum states, as well as by quantum blockchain proposals that make explicit use of time-entanglement for authentication and tamper detection\cite{PRA2025}.
Taken as a whole, the use of high-dimensional quantum encoding in combination
with time-entanglement provides a coherent framework for constructing
quantum-secure blockchain architectures. In the proposed protocol, security
properties such as authentication, data integrity, and non-repudiation are
derived from quantum-mechanical correlations rather than from assumptions of
computational hardness. Experimental and theoretical studies on
high-dimensional and time-entangled quantum systems indicate that the
required resources are increasingly accessible, supporting the view that the
present approach constitutes a realistic step toward implementable quantum
blockchain schemes in future quantum network environments \cite{PRA2025,OE2023}.

%
%

\ack{This study was produced from a portion of the first author's doctoral thesis under the supervision of the second author.}

\funding{This research received no external funding.}

\roles{\textbf{Arzu Akta\c{s}: }Conceptualization, Formal analysis, Investigation, Methodology, Validation, Visualization, Writing-original draft. \textbf{\.Ihsan Y\i lmaz:} Conceptualization, Formal analysis, Methodology, Validation, Supervision, Writing-review \& editing}

\bibliographystyle{vancouver}
\bibliography{sample}

\end{document}